# Towards the development of Dynamic Networked Psychology Hypotheses

**Liaquat Hossain**

Nebraska Healthcare Collaborative Chair of Population Health, Department Chairperson, Department of Cyber Systems, College of Business and Technology, The University of Nebraska at Kearney, NE 68849, Email: hossainl2@unk.edu

**Abstract**
One of the areas which is critical to bring back the social order is how individuals and communities show ability to cope aftermath of the disaster. It is now recognized that members of affected community in disasters prone areas are the first emergency responders. Individual and community psychology plays an important role in disaster management as human behaviour exhibit diverse risk perceptions, recognition of the threats that exists, positive and negative emotions, panic, anger, rumour, stress and learned helplessness. These psychological factors are important as lack of attention to these can lead to detrimental outcome of disaster management efforts. Disaster psychology has been seen as an emerging area of research and practice which deals with understanding of the psychological impact of individuals and community aftermath of the disasters. The aim of this paper is to put forward the conceptualization and development of dynamic networked psychology or DNP as a theoretical framework and its implications in exploring emotional contagion during disasters. We advocate theories of structural network dynamics can be used to construct DNP for exploring individuals as well as community's coping mechanisms for improving preparedness, response and recovery of disasters. The advent of computational social science or CSS promotes the empirical modelling and analysis of massive volume of user generated data by inferring meaningful patterns for finding answers to important social and behavioural science research questions dealing with individual and community coping ability. In presenting our DNP as a theoretical framework, we suggest that the underlying assumptions and integration of theories of social influence can be used to explore networks of emotional contagion for disasters. We conclude that the implications of our DNP framework are two folds: (1) analysis and predicting of the coping ability of social system relating to disasters; and, (2) a set of measures with underlying hypotheses for DNP as conceptual and theoretical framework in empirical modelling of networks of emotional contagion for disasters.

**Introduction**
Disasters may come from sources such as failure of physical systems, natural, man-made deliberate attack creating impact on the societies at multi level. Our social, physical and cyber systems are increasing becoming vulnerable to aforementioned disasters. It is therefore paramount to understand the intersection of physical, life and social systems in exploring how to build resilience for communities. There is also wider recognition that disaster management activities related to early warnings, detection, mitigation plan, and response requires a holistic approach to bring back the functioning of societies. Critical to the development and sustainability of community resilience is to develop support mechanisms for assisting communities to develop better coping ability aftermath of the disasters.

Research in mathematical and economic sociology (Galam, Gefen and Shapir, 1982; Granovetter, 1985), social anthropology (Wolfe, 1979), community and social psychology (Rappaport, 1987; Marsden and Friedkin, 1993) and computational psychology (Boden and Mellor, 1984) have long been advocating that social connectedness have significant impact on individual behaviour. Network studies have further provided us with the recognition that social phenomena are less about the behavior of individuals. It is rather regarded as the collections of individuals in groups, crowds, organizations, markets, classes, and even entire societies, all of which interact with each other via networks of information and influence, which in turn change over time (Watts and Strogatz, 1998). There have been significant interests among social psychologist, mathematical sociologist and information science research community to work towards the construction of the newly emerging area referred as computational social science or CSS.

Computing revolution of the past two decades with increased speed and memory for collecting and analyzing large scale social data has the potential to revolutionize traditional social science, leading to a new paradigm of "computational social science" (Lazer, Pentland, Adamic, Aral et al., 2009). CSS provides modelling approaches of user generated interactions and content in traditional community or virtual settings for exploring evolution, dynamical changes and adaptation of human behaviour. Integration of data sources from multiple sources in a single predictive framework can provide us with the avenue for conducting longitudinal observation of how individuals and inter-connected community members exchange information and how these interactions drive societal outcome. The functional characteristics of social media such as users' active and passive collective participation in providing opinion, referral and knowledge exchange has significant impact on disaster recovery process. It can therefore aid us in exploring their exchange patterns for conducting empirical investigation in supporting the idea of spread of emotions via social networks. We have recently seen evidence of large-scale experimental study using social media data for exploring how emotions both positive and negative can propagate through users social networks suggesting the notion of social contagion (Kramer, Guillory and Hancock, 2014).

Three important questions surface in our paper are: (1) How can we leverage CSS in exploring coping mechanisms of people in affected areas using real-time user generated content and interactions both online and network surveys?, (2) If early warnings are important for detection, devising mitigation plan as well as timely response, then, what is the impact of DNP as a theoretical framework, which can be used as analytical and predictive tool for studying the behavioural dynamics of individual and groups?, and, (3) How can the conceptual, theoretical and empirical modelling approach proposed in our DNP for exploring emotional contagion for disasters help facilitate building and sustaining resilience? In this paper, we focus our effort on the development of dynamic networked psychology or DNP as a theoretical framework through integration of the concepts of structural cohesion, structural equivalence and dynamic networks. Secondly, we explore the use of DNP as analytical and predictive tool for studying the behavioural dynamics of individual and groups. Thirdly, we suggest how DNP can guide us in exploring emotional contagion during disasters for building and sustaining resilience. Fourthly, we discuss the implications of our work in analysing and predicting the coping ability of social system relating to disasters as well as proposing a set of measures with underlying hypotheses or propositions for DNP in empirical modelling of networks of emotional contagion during disasters. Our paper is organised in six sections leading to the development and implications of DNP. In section two, we provide the conceptually clarification of networks in psychology. We then discuss computational social science in studying human behaviour in section three. Section four presents the role of psychology in disasters. In section five, we propose the development of dynamic networked psychology or DNP as a theoretical framework together with the implications of our proposed DNP for Disasters. Lastly, overall conclusion of our work is presented in section six of this paper.

**Concept of Networks in Psychology**
Psychology can be referred as the science of human behaviour, while social psychology advocates the study of human interactions (Gergen, 1973). Exploring human interactions from the perspective of networks can be seen as the social constructivist paradigm to explore how social forces, discourse and contextual artefacts help shaping the self or individual behaviour (Gergen, 1973; 1985; Cushman, 1990). This suggests the notion that self-structure constructed via social interactions have a tendency to change over time promoting the idea of psychology as dynamical complex systems (Vallacher and Nowak, 1994; Vallacher, Read and Nowak, 2002; Balas-Timar, 2014). Research is sociology, social psychology and computational social science promotes the idea of social connectedness having significant impact on individual behaviour in which interactions via networks of information and influence help shape the behaviour of individuals embedded within social structures (Watts and Strogatz, 1998). Empirical modeling of change in behavior as dynamical systems through social interactions and connectedness representing the notion of social contagion has been documented using massive online user interactions data have been documented in recent studies (Kramer, Guillory and Hancock, 2014). This suggests the notion of dynamic psychology, which aims to investigate the complex relationships of individuals and their environment for understanding how behavioural outcome is guided by different social forces (Rapaport, 1966).

Study of human interactions using networks perspective provides with generalized theoretical propositions for the study of human behaviour and their interactions. It further offers us with network-based measurements and observational framework to empirically model human behaviour, interactions and changes as dynamical complex systems (Schmittmann, Cramer, Waldorp, Epskamp, Kievit and Borsboom, 2013). Using networks perspective to study links and relationships linking different levels of analysis with its impact in driving social change such as health education and promotion have been well documented in community psychology (Sarason, 1976; Neal and Christens, 2014). This approach to studying social influence and its effect on individual behaviour as a dynamical process moderated by link formation and interactions has received significant attention in social psychology research (Mason, Conrey and Smith, 2007).

Therefore, networks approach to investigating behaviour as dynamical process allows us to explore how changes in behaviour occurs to social influence through formation of links and interactions (Kenrick, Li and Butner, 2003; Steglich, Snijders and Pearson, 2010). For example, social influence on spread of happiness (Fowler and Christakis, 2008), smoking behaviour (Christakis and Fowler, 2008; Mercken, Steglich, Sinclair, Holliday and Moore, 2012), spread of obesity (Christakis and Fowler, 2007), spread of alcohol consumption (Rosenquist, Murabito, Fowler and Christakis, 2010) to name a few using longitudinal network observation in health psychology and sociology provides us with empirical account into how social influences through links formation and interactions have impact of individual behaviour. However, the modelling of social links and interactions at multi-level using longitudinal network analysis requires us to explore the newly emerging area referred as computational social science for studying human behaviour.

**Computational social science in studying human behaviour**
The emergence of computational social science or CSS attributes to the availability of massive volume of user generated interactions and content in social media. CSS provides us with theoretical and modelling approaches to develop quantitative understanding of complex social systems dealing with social, behavioural and organizational science research questions (Lazer, Pentland et al., 2009; Conte, Gilbert, Bonelli, Cioffi-Revilla et al., 2012). This development provides advancement for social scientist traditionally investigating human behavior based on one time snapshot with small scale sampled observational data. It offers us with a set of generalized principles and modeling approaches to explore sociological network theories which in its conception lack large scale temporal behavioral interactions data. In particular, CSS has been seen as a model-based science for both explanatory and predictive modeling of social systems in which we can explore the emergent properties of social behavior; aggregate social behavior; and, institutional emergence (Conte, Gilbert et al., 2012).

Emerging properties of social behavior can be explored from the perspectives of self-organizing systems in which centralized control is not required for the rise of orderly patterns (Resnick, 1996). By examining the behaviour of individual within the context of global network, we can explore the behaviour of a local agent in comparison to global structure for understanding the structural changes of local nodes or a group of interconnected nodes are influenced by their interactions with the global network structure. Dynamic psychology advocates that social behaviour and its changes through adaptation process is essential spatial and temporal. In computer mediated systems users of different on-line community interact and interconnect with each other regardless of geographical and time distance. The social behaviour that we are observing in on-line communities presents a very dynamic, elastic and complex interactions representative of large social networks, in which individual behaviour can be conceived as part of their social belongings within the network that they subscribe.

The study of aggregate social behavior as emergent property was influenced by Schelling's empirical observation of segregation model providing the basis for understanding of group formation (Schelling, 1969; 1971). Segregation can be occurred through organized set of activities, economically determined, can result from communication as well as discriminatory choices of individuals. Advances in CSS-Computational Social Science and dynamic network modelling provides us with necessary tools for exploring dynamic group formation and emergent behaviour at an aggregate level through investigation of different network structures and their properties they exhibit during the emergence process. In doing so, we may also predict the emergence of social structures dynamically exhibiting different properties of networks which can be correlated against different outcome. Similarly, we can also model different emergent institutional forms. Computational science can be applied to real world scenario in developing quantitative and qualitative understanding of complex social phenomenon capturing adaptive human behaviour in dealing with uncertainty (Conte, Gilbert et al., 2012). Use of CSS can facilitate the real time analysis of massive volume of open data for understanding user behaviour and the dynamics of changes of behaviour for complex social phenomenon. This has significant implications for understanding psychology in disasters which may provide insights into community resilience and further aid our attempt to improve different aspects of disaster management.

**Psychology in Disasters**
Individuals and communities show ability to cope aftermath of the disaster is considered critical to bring back the social order aftermath of the disaster. Psychology in disasters deserves importance due to disaster survivor driven effort of fist search and rescue efforts aftermath of the disasters (Perry and Lindell, 2003). For example, studies of disaster management for 1979 tornado in Wichita Fall, TX suggest that only 13% percent of 5000 victims were rescued by the emergency management officials (Gantt and Gantt, 2012). We have also observed during 1985 Mexico City earthquake that the disaster management effort was supported by 2.8 million volunteers to took part in the response (Heide, 2004). The study of social and emotional contagion through networks during disasters is important for understanding community coping mechanisms as well as to develop effective disaster management activities to support the community well-being in affected areas. There is increasing interest in network studies to explore emotional contagion as an outcome of social influence, however, little attention has been given within the context of disasters to explore how user generated interactions and content can be leveraged to explore the dynamics of emotional contagion during disasters in an attempt to support community resilience.

Emotional contagion refers to the transfer of positive or negative emotions to others where people experience similar emotional states through their interactions within a given social network (Kramer, Guillory, Hancock, 2014). Experimental studies in laboratory setting provided us with empirical account into the transfer of positive and negative emotions to others resulted in emotional contagion (Hatfield, Cacioppo, Rapson, 1993). Emotional contagion is an observed social phenomenon in which individuals' positive (i.e., happiness, love) or negative (i.e., fear, anger and sadness) emotion can be influenced by others who are surrounding neighbours (Hatfield, Cacioppo, Rapson, 1993; Tsai, Bowring, Marsella, Tambe, 2011). Darwin's philosophy of human evolution documented the contagious notion of emotions as part of social and environmental systems (Doherty, 1997). However, measuring the likelihood of being contagious of emotions of others present significant modelling challenges. In this regard, Doherty (1997) suggested emotional states as happiness, love, fear, anger and sadness and further proposed fifteen items scale to measure the different emotional states. Items such as being with a happy person, gestures of someone showing positive attitudes making one feel warm and being around happy people influencing ones positive state has been used to measure the emotional contagion scale of happiness. Items relating to someone's expression of negative emotion having an impact on one's emotion have been used to measure the emotional contagion scale of sadness. These suggest that different aspect of network dynamics have positive and negative emotional influences on individuals who are part of a given network structure suggesting the likelihood of propagation of emotional contagion. In this regard, network theory such as social influence theory and its underlying assumptions provide us with the theoretical foundations to explore the emotional contagion.

Contagion referred as influence, which can be explored using the underlying theoretical assumptions drawn from social influence theory. Social influence theory has been used to explore the temporal relationships between behaviour of individuals at a given time with interactions in a neighbourhood network leading to suggesting direct influence of neighbour on behavioural outcome (Shalizi and Thomas, 2011). It has been used as theoretical construct for liking the structure of social relations to attitudes and behaviours of individuals who are part of a network (Marsden and Friedkin, 1993). The social influence theory assumes that the influence of actors on behavioural outcome is based on the proximity of the actors within a given network, which has been empirically studied in social contagion and innovation diffusion (Burt, 1987). It aims to explore three important attributes relating to the processes of social influence: (1) structural effects on behaviour of actors; (2) proximity of actors within a network and its effect on behaviour; and, (3) predictive success using mathematical and statistical modelling approaches.

Structural cohesion and equivalence are regarded as two prescribed approaches for exploring proximity in social influence processes (Marsden and Friedkin, 1993). Structural cohesion assumes the existence of network connectivity in terms of the number, length and strengths of the paths among actors. With the context of spread of innovation, the importance of physical proximity between ego and alters on the adoption process of innovation has been well documented (Burt, 1987). It has been found that there is a natural tendency to influence alters adoption to ego by creating awareness, suggesting issues related to trail as well as the consequences of adoption for alters. Therefore, the socialization of ego and alter forms the basis for cohesion model in which the frequency of communication between ego and alter can be used as a way to explore how alter's adoption can trigger the interest of ego towards adoption of innovation. Structural equivalence, on the other hand, focuses on the competitive nature of ego and alter where they may have identical relations with all other individuals within a network (Burt, 1987). It is concerned with equivalence in network positions among actors in a given network as a measure of proximity (Lorrain and White, 1971). In conclusion, contagion in a network structure can be predicted by both cohesion and equivalence for individuals who are tied to each other and individuals who are tied to other persons. Below, we present our DNP theoretical framework and propose a set of hypotheses with discussions.

**Dynamic Networked Psychology**
Figure 1 below presents dynamic networked psychology as a theoretical framework in which underlying assumptions and integration of social influence theory provides us with the basis for exploring networks of emotional contagion during disasters. In our conceptualization of DNP in figure 1, we suggest that social influence theory and its underlying assumptions can be used to explore disaster emotional contagion at aggregate, social behaviour and institutional emergence level through empirical modelling and analysis approaches drawn from computational social science. Below, we propose and explain a number of hypotheses that emerges from our DNP conceptualization linking social influence theory, disaster emotional contagion and computational social science.

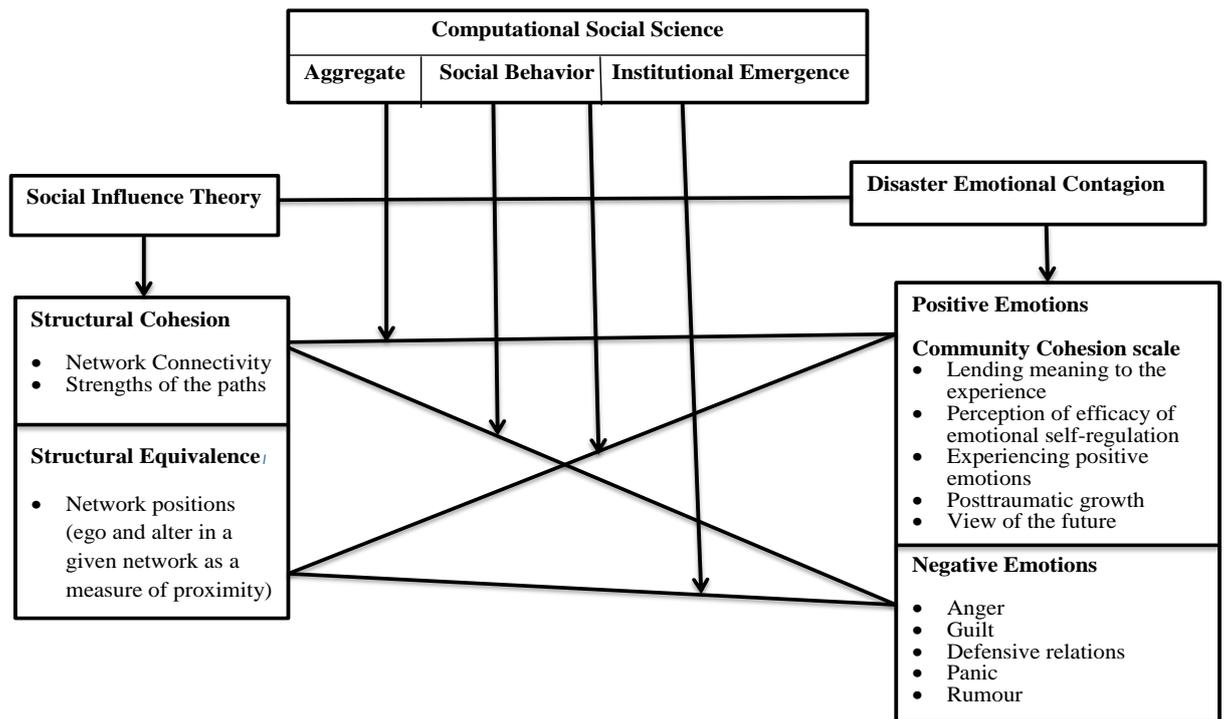

**Figure 1. Dynamic networked psychology as a theoretical framework**

Structural cohesion models provide explanatory models showing how individuals' attitudes and behaviours are the outcome of cohesion (Friedkin, 1984; Moody and White, 2003). It is defined as a structure in which the group can be disconnected if minimum numbers of actors are removed from the structure (Moody and White, 2003). Cohesion can further suggest how well the group can hold together and function under disruption (Gross and Martin, 1952). The relationship between structural cohesion and emotional contagion can be explored if we assume that a person or group can potentially influence the emotions of behaviour of other person or group through social influence which have been documented in organizational, social and community psychology research (Cacioppo and Petty, 1987; Schoenewolf, 1990; Levy and Nail, 1993; Barsade, 2002).

Structural equivalence models are not regarded as explanatory models (Friedkin, 1984), which can be observed in a given network where members are characterized as structurally equivalent if they share relation to other members or types of members (Lorrain and White, 1971; Sailer, 1978). It suggests social influence as the outcome of social position where structural equivalence is observed when individuals share network ties (Burt, 1993; Lorrain and White, 1971) and is regarded as an alternative to structural cohesion model (Burt, 1987; Wasserman and Faust, 1994). Structural equivalence hypothesis relates to the exploring whether network positions such ego and alter in a given network as a measure of proximity can be used to explain positive and negative emotions at aggregate, social behaviour and institutional emergence level. In our DNP theoretical framework presented above, we highlight the measurement of disaster emotional contagion based on five items of community cohesion scale derived from Doherty's (1997) validated emotional contagion scale, which has been applied with the context of disasters to explore emotional contagion aftermath of disasters (Pérez-Sales, et. al., 2005; Vázquez, et. al., 2005). Our structural cohesion and structural equivalence hypotheses related to disaster emotional contagion are presented below:

H1: Network connectivity such as number of links and lengths correlates to positive or negative emotions at aggregate, social behaviour and institutional emergence level.

H2: Strengths of the paths correlate to positive or negative emotions at aggregate, social behaviour and institutional emergence level.

H3: Network positions such ego and alters in a given network as a measure of proximity correlates to positive and negative emotions at aggregate, social behaviour and institutional emergence level.

The implications of our proposed dynamic networked psychology framework are two folds: (1) scalability in analysis and prediction of coping ability of social system for disasters; and, (2) a set of measures with underlying hypotheses of networked psychology as conceptual and theoretical framework in empirical modelling of networks of emotional contagion for disasters. We highlight that the advances in computational social science of CSS offers us opportunity to observe, construct and predict different types of emotional contagion such as coping mechanisms of people through social influence in affected areas. Using real-time user generated content and interactions both online and network surveys provide us with an avenue in addressing scalability for analysing and predicting the coping ability of social system. Community-based analysis highlight that condition such as community awareness, coping ability, previous experience can shape exposures, adaptive capacities for supporting community specific needs (Smit and Wandel, 2006). It further provides us with the opportunity for detecting early warnings, devising mitigation plan as well as timely response to disasters. Therefore, DNP as a theoretical framework can be used as analytical and predictive tool for studying the behavioural dynamics of individual and groups for exploring disaster emotional contagion, which in turn can help to facilitate building and sustaining resilience.

**Conclusions**
Affected community members in disasters prone areas are increasingly regarded as the first emergency responders. In this paper, we suggest that understanding of individuals and communities coping aftermath of the disaster is critical to bring back the social order. This deserves importance as human behaviour for disasters exhibit a diverse set of issues ranging from risk perceptions, positive and negative emotions to panic, rumour and stress. Lack of attention to these can lead to negative outcome in disaster management efforts. To address these challenges, the focus on disaster psychology for understanding of the psychological impact of individuals and community aftermath of the disasters is critical. Our proposed dynamic networked psychology as a theoretical framework using both online and network surveys can be used as analytical and predictive tool for studying the behavioural dynamics of individual and groups for disasters. In particular, the conceptual, theoretical and empirical modelling approach of DNP provides us with a set of hypotheses related emotional contagion for disasters. Our proposed DNP construction is based on the theories of structural dynamics of networks for exploring individuals as well as community's coping mechanisms for disasters. We conclude that the implications of our DNP framework are two folds: (1) analysis and predicting of the coping ability of social system relating to disasters; and, (2) a set of measures with underlying hypotheses for DNP as conceptual and theoretical framework in empirical modelling of networks of emotional contagion for disasters.

**References**

1. Aguirre, B. E. (2005). Emergency evacuations, panic, and social psychology. *Psychiatry: Interpersonal and Biological Processes*, *68*(2), 121-129.
2. Aylward, B. S., Odar, C. C., Kessler, E. D., Canter, K. S., & Roberts, M. C. (2012). Six degrees of separation: An exploratory network analysis of mentoring relationships in pediatric psychology. *Journal of Pediatric psychology*, *37*(9), 972-979.
3. Balas-Timar, D. (2014). Is it psychology about linear or dynamic systems. *SEA-Practical Application of Science*, *2*(2), 4.
4. Barling, J., Bluen, S. D., & Fain, R. (1987). Psychological functioning following an acute disaster. *Journal of Applied Psychology*, *72*(4), 683.
5. Barsade, S. G. (2002). The ripple effect: Emotional contagion and its influence on group behavior. *Administrative Science Quarterly*, *47*(4), 644-675.
6. Boden, M. A., & Mellor, D. H. (1984). What is computational psychology?.*Proceedings of the Aristotelian Society, Supplementary Volumes*, *58*, 17-53.
7. Bonanno, G. A., Galea, S., Bucciarelli, A., & Vlahov, D. (2007). What predicts psychological resilience after disaster? The role of demographics, resources, and life stress. *Journal of consulting and clinical psychology*, *75*(5), 671.
8. Bosse, T., Hoogendoorn, M., Klein, M. C., Treur, J., Van Der Wal, C. N., & Van Wissen, A. (2013). Modelling collective decision making in groups and crowds: Integrating social contagion and interacting emotions, beliefs and intentions. *Autonomous Agents and Multi-Agent Systems*, 1-33.
9. Burt, R. S. (1987). Social contagion and innovation: Cohesion versus structural equivalence. *American journal of Sociology*, *92*(6), 1287-1335.
10. Burt, R. S. (1993). The social structure of competition. *Explorations in economic sociology*, *65*, 103.
11. Cacioppo, J. T., & Petty, R. E. (1987). Stalking rudimentary processes of social influence: A psychophysiological approach. In *Social Influence: The Ontario Symposium: Ontario Symposium on Personality and Social Psychology* (Vol. 5, pp. 41-74). Hillsdale, NJ: Erlbaum.
12. Christakis, N. A., & Fowler, J. H. (2007). The spread of obesity in a large social network over 32 years. *New England journal of medicine*, *357*(4), 370-379.
13. Christakis, N. A., & Fowler, J. H. (2013). Social contagion theory: examining dynamic social networks and human behavior. *Statistics in medicine*, *32*(4), 556-577.
14. Christakis, N. A., & Fowler, J. H. (2008). The collective dynamics of smoking in a large social network. *New England journal of medicine*, *358*(21), 2249-2258.
15. Conte, R., Gilbert, N., Bonelli, G., Cioffi-Revilla, C., Deffuant, G., Kertesz, J., ... & Nowak, A. (2012). Manifesto of computational social science.*The European Physical Journal Special Topics*, *214*(1), 325-346.
16. Cushman, P. (1990). Why the self is empty: Toward a historically situated psychology. *American psychologist*, *45*(5), 599.
17. Doherty, R. W. (1997). The emotional contagion scale: A measure of individual differences. *Journal of nonverbal Behavior*, *21*(2), 131-154.
18. Fowler, J. H., & Christakis, N. A. (2008). Dynamic spread of happiness in a large social network: longitudinal analysis over 20 years in the Framingham Heart Study. *Bmj*, *337*, a2338.
19. Friedkin, N. E. (1984). Structural cohesion and equivalence explanations of social homogeneity. *Sociological Methods & Research*, *12*(3), 235-261.
20. Galam, S., Gefen, Y., & Shapir, Y. (1982). Sociophysics: A new approach of sociological collective behaviour. I. mean-behaviour description of a strike. *Journal of Mathematical Sociology*, *9*(1), 1-13.

21. Gantt, P., & Gantt, R. (2012). Disaster psychology: dispelling the myths of panic. *Professional Safety*, *57*(08), 42-49.
22. Gergen, K. J. (1985). The social constructionist movement in modern psychology. *American psychologist*, *40*(3), 266.
23. Gergen, K. J. (1973). Social psychology as history. *Journal of personality and social psychology*, *26*(2), 309.
24. Granovetter, M. (1978). Threshold models of collective behavior. *American journal of sociology*, *83*(6), 1420-1443.
25. Granovetter, M. (1985). Economic action and social structure: The problem of embeddedness. *American journal of sociology*, *91*(3), 481-510.
26. Gross, N., & Martin, W. E. (1952). On group cohesiveness. *American Journal of Sociology*, *57*(6), 546-564.
27. Guastello, S. J., Koopmans, M., & Pincus, D. (2009). Chaos and complexity in psychology. *Cambridge, Cambridge University*.
28. Hatfield, E., Cacioppo, J. T., & Rapson, R. L. (1993). Emotional contagion. *Current directions in psychological science*, *2*(3), 96-100.
29. Heide, E. A. (2004). Common misconceptions about disasters: Panic, the "disaster syndrome," and looting. *The first 72 hours: a community approach to disaster preparedness*, *337*.
30. Hill, A. L., Rand, D. G., Nowak, M. A., & Christakis, N. A. (2010). Emotions as infectious diseases in a large social network: the SISa model. *Proceedings of the Royal Society of London B: Biological Sciences*, *277*(1701), 3827-3835.
31. Hobfoll, S. E., & Walfisch, S. (1984). Coping with a threat to life: A longitudinal study of self-concept, social support, and psychological distress. *American journal of community psychology*, *12*(1), 87-100.
32. Jason, L. A., Light, J. M., Stevens, E. B., & Beers, K. (2014). Dynamic social networks in recovery homes. *American journal of community psychology*, *53*(3-4), 324-334.
33. Kramer, A. D., Guillory, J. E., & Hancock, J. T. (2014). Experimental evidence of massive-scale emotional contagion through social networks. *Proceedings of the National Academy of Sciences*, *111*(24), 8788-8790.
34. Kenrick, D. T., Li, N. P., & Butner, J. (2003). Dynamical evolutionary psychology: individual decision rules and emergent social norms. *Psychological review*, *110*(1), 3.
35. Kossinets, G., & Watts, D. J. (2009). Origins of homophily in an evolving social network 1. *American journal of sociology*, *115*(2), 405-450.
36. Lazer, D., Pentland, A. S., Adamic, L., Aral, S., Barabasi, A. L., Brewer, D., ... & Jebara, T. (2009). Life in the network: the coming age of computational social science. *Science, 323*(5915), 721.
37. Levy, D. A., & Nail, P. R. (1993). Contagion: A theoretical and empirical review and reconceptualization. *Genetic, social, and general psychology monographs*, 119, 233-284.
38. Lorrain, F., & White, H. C. (1971). Structural equivalence of individuals in social networks. *The Journal of mathematical sociology*, *1*(1), 49-80.
39. Marsden, P. V., & Friedkin, N. E. (1993). Network studies of social influence. *Sociological Methods & Research*, *22*(1), 127-151.
40. Mason, W. A., Conrey, F. R., & Smith, E. R. (2007). Situating social influence processes: Dynamic, multidirectional flows of influence within social networks. *Personality and social psychology review*, *11*(3), 279-300.
41. Mercken, L., Steglich, C., Sinclair, P., Holliday, J., & Moore, L. (2012). A longitudinal social network analysis of peer influence, peer selection, and smoking behavior among adolescents in British schools. *Health Psychology*, *31*(4), 450.
42. Moody, J., & White, D. R. (2003). Structural cohesion and embeddedness: A hierarchical concept of social groups. *American Sociological Review*, 103-127.
43. Neal, J. W., & Christens, B. D. (2014). Linking the levels: Network and relational perspectives for community psychology. *American journal of community psychology*, *53*(3-4), 314-323.

44. Pérez-Sales, P., Cervellón, P., Vázquez, C., Vidales, D., & Gaborit, M. (2005). Post-traumatic factors and resilience: the role of shelter management and survivours' attitudes after the earthquakes in El Salvador (2001). *Journal of Community & Applied Social Psychology*, *15*(5), 368-382.
45. Perry, R. W., & Lindell, M. K. (2003). Preparedness for emergency response: guidelines for the emergency planning process. *Disasters*, *27*(4), 336-350.
46. Raafat, R. M., Chater, N., & Frith, C. (2009). Herding in humans. *Trends in cognitive sciences*, *13*(10), 420-428.
47. Rand, D. G., Arbesman, S., & Christakis, N. A. (2011). Dynamic social networks promote cooperation in experiments with humans. *Proceedings of the National Academy of Sciences*, *108*(48), 19193-19198.
48. Rapaport, D. (1966). Dynamic psychology and Kantian epistemology. *Journal of the History of the Behavioral Sciences*, *2*(3), 192-199.
49. Rappaport, J. (1987). Terms of empowerment/exemplars of prevention: Toward a theory for community psychology. *American journal of community psychology*, *15*(2), 121-148.
50. Resnick, M. (1996). Beyond the centralized mindset. *The journal of the learning sciences*, *5*(1), 1-22.
51. Rosenquist, J. N., Murabito, J., Fowler, J. H., & Christakis, N. A. (2010). The spread of alcohol consumption behavior in a large social network. *Annals of internal medicine*, *152*(7), 426-433.
52. Sailer, L. D. (1978). Structural equivalence: Meaning and definition, computation and application. *Social Networks*, *1*(1), 73-90.
53. Sarason, S. B. (1976). Community psychology, networks, and Mr. Everyman. *American Psychologist*, *31*(5), 317.
54. Schelling, T. C. (1969). Models of segregation. *The American Economic Review*, *59*(2), 488-493.
55. Schelling, T. C. (1971). Dynamic models of segregation. *Journal of mathematical sociology*, *1*(2), 143-186.
56. Shalizi, C. R., & Thomas, A. C. (2011). Homophily and contagion are generically confounded in observational social network studies. *Sociological methods & research*, *40*(2), 211-239.
57. Sharpanskykh, A., & Treur, J. (2014). Modelling and analysis of social contagion in dynamic networks. *Neurocomputing*, *146*, 140-150.
58. Squazzoni, F. (2008). The micro-macro link in social simulation. *Sociologica*, *2*(1), 0-0.
59. Schweers Cook, K. (2005). Networks, norms, and trust: The social psychology of social capital∗ 2004 Cooley Mead Award Address. *Social Psychology Quarterly*, *68*(1), 4-14.
60. Schmittmann, V. D., Cramer, A. O., Waldorp, L. J., Epskamp, S., Kievit, R. A., & Borsboom, D. (2013). Deconstructing the construct: A network perspective on psychological phenomena. *New ideas in psychology*, *31*(1), 43-53.
61. Schoenewolf, G. (1990). Emotional contagion: Behavioral induction in individuals and groups. *Modern Psychoanalysis*, 15, 49-61.
62. Skyrms, B., & Pemantle, R. (2000). A dynamic model of social network formation. *Proceedings of the national academy of sciences*, *97*(16), 9340-9346.
63. Smith, E. R. (1996). What do connectionism and social psychology offer each other?. *Journal of personality and social psychology*, *70*(5), 893.
64. Smit, B., & Wandel, J. (2006). Adaptation, adaptive capacity and vulnerability. *Global environmental change*, *16*(3), 282-292.
65. Steglich, C., Snijders, T. A., & Pearson, M. (2010). Dynamic networks and behavior: Separating selection from influence. *Sociological methodology*, *40*(1), 329-393.
66. Tsai, J., Bowring, E., Marsella, S., & Tambe, M. (2011, September). Empirical evaluation of computational emotional contagion models. In *International Workshop on Intelligent Virtual Agents* (pp. 384-397). Springer, Berlin, Heidelberg.

67. Vallacher, R. R., Read, S. J., & Nowak, A. (2002). The dynamical perspective in personality and social psychology. *Personality and Social Psychology Review*, *6*(4), 264-273.
68. Vázquez, C., Cervellón, P., Pérez-Sales, P., Vidales, D., & Gaborit, M. (2005). Positive emotions in earthquake survivors in El Salvador (2001).*Journal of Anxiety Disorders*, *19*(3), 313-328.
69. Vorst, H. C. (2010). Evacuation models and disaster psychology. *Procedia Engineering*, *3*, 15-21.
70. Wasserman, S., & Faust, K. (1994). *Social network analysis: Methods and applications* (Vol. 8). Cambridge university press.
71. Watts, D. J., & Strogatz, S. H. (1998). Collective dynamics of 'small-world'networks. *Nature*, *393*(6684), 440-442.
72. Westaby, J. D., Pfaff, D. L., & Redding, N. (2014). Psychology and social networks: a dynamic network theory perspective. *American Psychologist*, *69*(3), 269.
73. Wolfe, A. W. (1979). The rise of network thinking in anthropology. *Social Networks*, *1*(1), 53-64.